\documentclass[aps,prapplied,showpacs,twocolumn,preprintnumbers,amsmath,amssymb,longbibliography]{revtex4-2}
\usepackage{graphicx}
\usepackage{dcolumn}
\usepackage{bm}
\usepackage{color}
\usepackage{amsmath}
\usepackage{upgreek}
\usepackage{float}
\usepackage[colorlinks,linkcolor=blue, urlcolor=blue, anchorcolor=blue, citecolor=blue]{hyperref}

\begin{document}

\title{Antiferromagnetic spatial photonic Ising machine through optoelectronic correlation
computing}

\author{Junyi Huang}
\thanks{These authors contributed equally to this work.}
\affiliation{Interdisciplinary Center of Quantum Information, State Key Laboratory of Modern Optical Instrumentation, and Zhejiang Province Key Laboratory of Quantum Technology and Device, Department of Physics, Zhejiang University, Hangzhou 310027, China}

\author{Yisheng Fang}
\thanks{These authors contributed equally to this work.}
\affiliation{Interdisciplinary Center of Quantum Information, State Key Laboratory of Modern Optical Instrumentation, and Zhejiang Province Key Laboratory of Quantum Technology and Device, Department of Physics, Zhejiang University, Hangzhou 310027, China}

\author{Zhichao Ruan}
\email{zhichao@zju.edu.cn}
\affiliation{Interdisciplinary Center of Quantum Information, State Key Laboratory of Modern Optical Instrumentation, and Zhejiang Province Key Laboratory of Quantum Technology and Device, Department of Physics, Zhejiang University, Hangzhou 310027, China}

\begin{abstract}
Recently, spatial photonic Ising machines (SPIM) have been demonstrated to compute the minima of Hamiltonians for large-scale spin systems. Here we propose to implement an antiferromagnetic model through optoelectronic correlation computing with SPIM. Also we exploit the gauge transformation which enables encoding the spins and the interaction strengths in a single phase-only spatial light modulator. With a simple setup, we experimentally show the ground state search of an antiferromagnetic model with $40000$ spins in number-partitioning problem. Thus such an optoelectronic computing exhibits great programmability and scalability for the practical applications of studying statistical systems and combinatorial optimization problems.
\end{abstract}
\maketitle

\section{Introduction}
The spin glass models are widely used for investigations of interacting systems in both science and engineering \cite{binder1986spin, edwards1975theory, sherrington1975solvable, gabay1981coexistence, reger1986three, de2009unifying, banuls2020simulating, hopfield1982neural, amit1985spin, agliari2013parallel,sourlas1989spin, nishimori2001statistical}. In the past decades, the developments in spin machines have generated tremendous interest due to the prospects of solving a large class of NP-hard problems by searching the ground states of spin system Hamiltonians \cite{lucas2014ising}. Optimization problem solvers with remarkable performance have been demonstrated in various systems, e.g. trapped ions \cite{ kim2010quantum,britton2012engineered,ma2011quantum}, atomic and photonic condensates \cite{struck2013engineering,kassenberg2020controllable}, superconducting circuits \cite{johnson2011quantum}, coupled parametric oscillators \cite{ mcmahon2016fully,inagaki2016coherent,inagaki2016large,bohm2019poor,hamerly2019experimental,bello2019persistent,bohm2018understanding,tiunov2019annealing,wang2013coherent, yamamoto2017coherent, marandi2014network}, injection-locked or degenerate cavity lasers \cite{utsunomiya2011mapping, takata2012transient, babaeian2019single, tradonsky2019rapid}, integrated nanophotonic circuits \cite{roques2020heuristic,prabhu2020accelerating,shen2017deep,wu2014optical,vazquez2018optical,okawachi2020nanophotonic}, and polaritons \cite{berloff2017realizing,kalinin2018simulating,kalinin2020polaritonic}.

Recently, like optical analog computations exploring the spatial degrees of freedom \cite{silva2014performing,bykov2014optical,ruan2015spatial,youssefi2016analog,zhu2017plasmonic,zhang2018implementing,guo2018photonic,zhu2019generalized,zangeneh2020analogue}, the spatial photonic Ising machine (SPIM) has been proposed with reliable large-scale Ising spin systems, even up to thousands of spins \cite{pierangeli2019large}. With spatial light modulations, these spatial Ising spin setups benefit from the high speed and parallelism of optical signal processing \cite{pierangeli2019large, pierangeli2020noise, pierangeli2020adiabatic, pierangeli2021scalable, kumar2020large}. Although the modeling of ferromagnetic and spin glass systems has been demonstrated \cite{fang2021experimental}, how to implement antiferromagnetic models in SPIM has not been proposed yet. In particular, the antiferromagnetic Ising models are important and extensive in research fields like oxide materials \cite{shull1951neutron} and giant magnetoresistance \cite{binasch1989enhanced,baibich1988giant}. Also, the combinatorial optimization problems with antiferromagnetic Ising models have many real-world applications such as multiprocessor scheduling, minimization of circuit size and delay, cryptography, and logistics analysis \cite{grass2016quantum,mertens2006easiest}.

In this Work, we propose to implement the antiferromagnetic model through optoelectronic correlation computing. We show that an antiferromagnetic Hamiltonian can be evaluated through the correlation between a distribution function and the measured optical intensity with SPIM. We experimentally demonstrate the ground-state-search process with a number-partitioning problem, where $40000$ spins are connected with random antiferromagnetic interaction strengths. Our results show that the proposed antiferromagnetic model in SPIM can evolve toward the ground state, exhibiting an efficient approach by scalable degrees of freedom in spatial light modulation.
\section{Results}

\subsection{ Optoelectronic correlation computing for Mattis-type Ising model}

\begin{figure*}[htb]
\includegraphics[width=120mm]{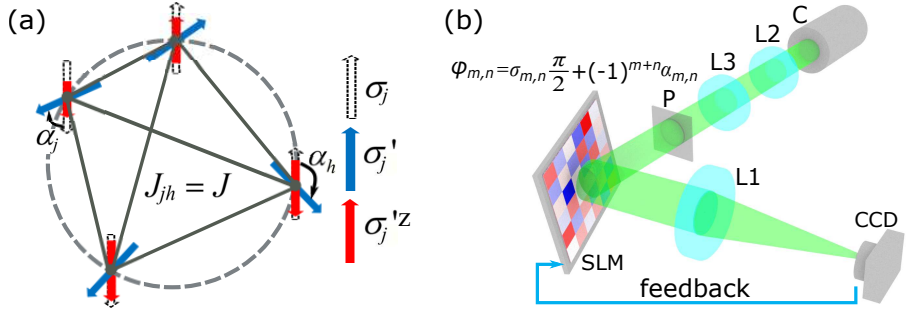}
\caption{ \label{fig:1} (a) Schematic of the gauge transformation. The dotted, blue and red arrow correspond to the origin spin ${{\sigma }_{j}}$, the new spin vector ${\sigma'_j}$ and the effective spin vector ${\sigma'_j}^{z}$, respectively. (b) The experimental optical setups for SPIM: SLM, spatial light modulator; C, collimator; L, lens; P, polarizer. A collimated laser beam illuminates the SLM. By gauge transformation, the gauge-transformed effective spin configuration ${{\mathbf{{S}'}}^{z}}=\left\{ {\sigma'_{m,n}}^{z} \right\}$ is encoded through a single phase-only SLM following Eq. (\ref{eq:1}). The modulated light is detected by a CCD camera in the back-focus plane of Lens L1.}
\end{figure*}

We first introduce the gauge transformation for SPIM, which encodes the spins and the interaction strengths in a single phase-only spatial light modulator (SLM) \cite{fang2021experimental}. We consider a Mattis-type spin glass system with the Ising model Hamiltonian $ H =- \sum\limits_{jh} J_{jh}\sigma _j\sigma _h$, where the spin configuration is $\mathbf{S}\text{=}\left\{ {{\sigma }_{j}} \right\}$ ($j=1,2,\cdots N$) and ${{\sigma }_{j}}$ takes binary value of either $+1$ or $-1$, representing the spin-up or spin-down state respectively. The interaction strengths can be expressed as ${{J}_{jh}}=J{{\xi }_{j}}{{\xi }_{h}}$, where $J$ is an interaction strength as a function of the distance between two spins with the unit of energy, and the amplitude modulation ${\xi }_{j}$ is limited as $-1\le {{\xi }_{j}}\le 1$. By the gauge transformation shown in Fig.~\ref{fig:1}(a), when rotating each original spin ${{\sigma }_{j}}$ with the angle ${{\alpha }_{j}}=\arccos {{\xi }_{j}}$, the new spin vector ${\sigma'_j}$ is projected on the $z$-axis to obtain the effective spin ${\sigma'_j}^{z}={{\xi }_{j}}{{\sigma }_{j}}$. As a result, the gauge transformation keeps the Hamiltonian invariant $ H =- \sum\limits_{jh} J{\sigma'_j}^z{\sigma'_h}^z$, while the interactions between the $z$ components of gauge-transformed spins are equal to the strength $J$.

The gauge invariance property promises that the experimental implementation only needs a single phase-only SLM, with uniform illumination by a collimated uniform laser beam [Fig.~\ref{fig:1}(b)]. This setup circumvents the difficulty of pixel alignment in the previously proposed SPIM \cite{pierangeli2019large, pierangeli2020noise, pierangeli2020adiabatic, pierangeli2021scalable}, and therefore greatly improves the system stability and the computing fidelity. Since the spins are loaded through two-dimensional spatial modulation, the $j$-th spin is distributed in a square lattice at $\mathbf{j}=(m,n)$, where $1\le m \le {{N}_{x}},\text{ }1\le n \le {{N}_{y}}$. According to Ref.\cite{fang2021experimental}, after the gauge transformation, each spin is encoded by a macropixel with phase modulation $\varphi_{m,n}$ such that ${{\sigma'_{j}}^z}=\exp(i \varphi_{m,n})$ and
\begin{equation}
\varphi_{m,n}=\sigma _{m,n}\frac{\pi }{2}+(-1)^{m+n}\alpha _{m,n} \label{eq:1}.
\end{equation}
Then with Lens L1 of the focal length $f$ performing Fourier transformation, we detect the band-limited intensity distribution $I(\mathbf{u})$ confined within the first diffraction order zone $A$ on the focal plane, and $I(\mathbf{u})=\sum\limits_{ {j} {h}} {{\sigma '_{j}}^z{\sigma '_{h}}^z} {e^{i\frac{{2\pi }}{{f\lambda }}({{\mathbf{x}}_ {j}} - {{\mathbf{x}}_ {h}}) \cdot {\mathbf{u}}}}{\text{sin}}{{\text{c}}^2}(\frac{{{\mathbf{u}}{{W}}}}{{f\lambda }})$, where $\lambda$ is the wavelength, $f$ is the focal length of lens L1, $W$ is the length of each macropixel, ${{\mathbf{x}}_{ {j}}}=W\mathbf{j}$ is the center position of the $j$-th pixel, $\mathbf{u}=(u,v)$ is the spatial coordinate in the focal plane, and $\text{sinc}(\mathbf{u})=\frac{\sin \pi {{u}}}{\pi {{u}}}\frac{\sin \pi {{v}}}{\pi {{v}}}$. Suppose that we preset a distribution function ${g_c}(\mathbf{u})$ and evaluate the correlation function $F$ as
\begin{equation}
{F} = \int\limits_{A} {I({\mathbf{u}}){g_c}({\mathbf{u}})} {\text{d}}{\mathbf{u}} = \sum\limits_{jh} {G({\mathbf{j}} - {\mathbf{h}}){\sigma'_j}^z{\sigma'_h}^z}
\label{eq:2}.
\end{equation}
Here
\begin{equation}
G({\mathbf{k}}) = \int\limits_{A} {{g_c}\left( {\mathbf{u}} \right) {\text{sin}}{{\text{c}}^2}(\frac{{{{W}}{\mathbf{u}}}}{{f\lambda}})}{e^{i\frac{{2\pi W}}{{f\lambda }}{\mathbf{k}} \cdot {\mathbf{u}}}}{\text{d}}{\mathbf{u}}
\label{eq:3},
\end{equation}
that is, ${G}({\mathbf{k}})$ is the Fourier transformation of ${g_c}\left( {\mathbf{u}} \right) {\text{sin}}{{\text{c}}^2}(\frac{{{{W}}{\mathbf{u}}}}{{f\lambda}})$. Indeed, Eq.~(\ref{eq:2}) shows that by presetting an appropriate ${g_c}$, a Mattis-type Ising Hamiltonian can be evaluated as
\begin{equation}
H= -F= -\sum\limits_{jh} {G({\mathbf{j}} - {\mathbf{h}}){\sigma'_j}^z{\sigma'_h}^z }.
\label{eq:4}
\end{equation}

We note that the distribution function $g_c({\mathbf{u}})$ is distinct from the target intensity $I_T({\mathbf{u}})$ proposed in Ref.~\cite{pierangeli2019large}. Here $g_c({\mathbf{u}})$ can be an arbitrary real function to guarantee an even function of the interaction strength [c.f. Eq.~(\ref{eq:3})], which has either positive or negative values, while the target intensity $ I_T$ always has non-negative values. In particular, for the antiferromagnetic model, all the interaction strengths ${{J}_{jh}}<0$. In the case that ${\xi }_{j}$s are positive, $G({\mathbf{j}}-{\mathbf{h}})$ should be negative to ensure antiferromagnetic interactions between all the spins. It leads that $g_c$ must be negative for some values of ${\mathbf{u}}$, otherwise Eq.~(\ref{eq:3}) shows $G>0$ for ${\mathbf{j}}={\mathbf{h}}$. Moreover, when $g_c$ has both positive and negative values, the Hamiltonian can be evaluated through the correlation function as Eq.~(\ref{eq:2}), while it cannot be implemented by the target-intensity approach.

\subsection{Number-partitioning problem with the antiferromagnetic Hamiltonian}

The antiferromagnetic model in SPIM provides a new computation platform for studying the challenging combinatorial optimization problems. As a demonstration, here we present the ground-state-search process of a combinatorial optimization problem, the NP-hard number-partitioning problem \cite{lucas2014ising, mertens1998phase}: One would like to divide a set $\Xi =\{\xi_j\}$, containing $N$ real numbers $(j=1,2,\cdots N)$, into two subsets $\Xi_1$ and $\Xi _2$, such that the difference between the summations of elements in two subsets $\sum _1=\sum\limits_{{\xi_j}\in {\Xi_1}}{\xi_j}$ and ${{\sum }_{2}}=\sum\limits_{{\xi_j}\in {\Xi_2}}{\xi_j}$ is as small as possible. Without loss of generality, we suppose all $\xi_j$s in the set $\Xi$ are real numbers belonging to the range $(0,1]$. Specifically, when the number set $\Xi$ has parity symmetry, the spins of such models can be analytically optimized. For instance, for the set with an even total number of $\xi_j$s, when $\xi_j = \xi_{N+1-j}$, the spin should be $\sigma_j = -\sigma_{N+1-j}$ to ensure the equivalence of two subsets. In general, by labeling the elements belonging to two different subsets $\Xi_1$ and $\Xi_2$ with $\sigma_j=1$ and $-1$ respectively, the optimization is equivalent to minimizing the antiferromagnetic Hamiltonian $H=(\sum\limits_{j}\xi_j\sigma_j)^2=\sum\limits_{jh}\xi_j\xi_h\sigma_j\sigma_h$.

To implement such an antiferromagnetic Hamiltonian in SPIM, we explore the gauge transformation to search a spin configuration $\mathbf{{S}'}=\left\{ {\sigma'_j}^{z} \right\}$ where ${\sigma'_j}^z=\xi_j\sigma_j$ while keeping the interaction strength between any two spins $ G({\mathbf{k}})=-1$. Due to the gauge invariance, the Hamiltonian is $H=\sum\limits_{jh}{\sigma'_j}^z{\sigma'_h}^z$ and the optimized value of $\left| {{\sum }_{1}}-{{\sum }_{2}} \right|$ is the total magnetization strength of the gauge-transformed spins $\left| {{m}'} \right|=\frac{1}{N}\left| \sum\limits_{j}{{\sigma'_j}^{z}} \right|$. During the experimental iterations, the spin configuration is updated gradually \cite{pierangeli2019large}, and the system definitely evolves to the ground states, indicating the process of solving the optimization problem.

\subsection{Experimental ground state search }

We experimentally demonstrate the ground-state-search process with the number-partitioning problem. As shown in Fig.~\ref{fig:1}(b), a collimated Gaussian beam (wavelength $\lambda =532\operatorname{nm}$) is expanded by two confocal lenses L2 ($50\text{mm}$ focal length) and L3 ($500\text{mm}$ focal length). After expansion, the waist radius of the collimated beam is about 36mm. Then a polarizer P is used to prepare the incident beam linearly polarized along the long display axis of the SLM (Holoeye PLUTO-NIR-011). The SLM is calibrated through the two-shot method based on generalized spatial differentiator \cite{huang2020two}. Lens L1 with the focal length $f=100\text{mm}$ performs Fourier transformation, where a CCD beam profiler (Ophir SP620) is used to detect the optical field intensity on the back focal plane.

\begin{figure}[htb]
\includegraphics[width=3.0in]{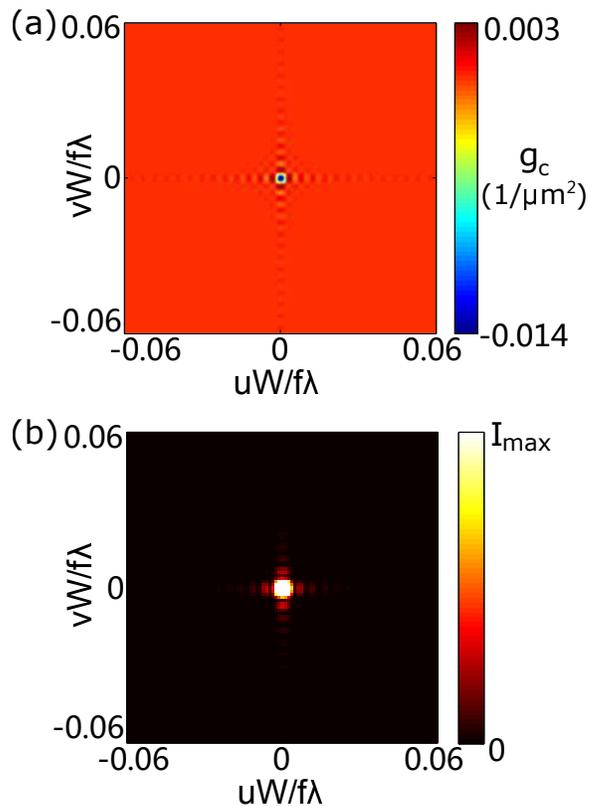}
\caption{\label{fig:2} (a) Calculated distribution function $g_c(\mathbf{u})$ of the antiferromagnetic Mattis model, where the interaction strength between any two spins $ G({\mathbf{k}})=-1$. (b) By setting the SLM with uniform phase modulation, we measure the intensity on CCD plane and thus determine the origin of coordinates at the maximal intensity. }
\end{figure}

\begin{figure*}[htb]
\includegraphics[width=160mm]{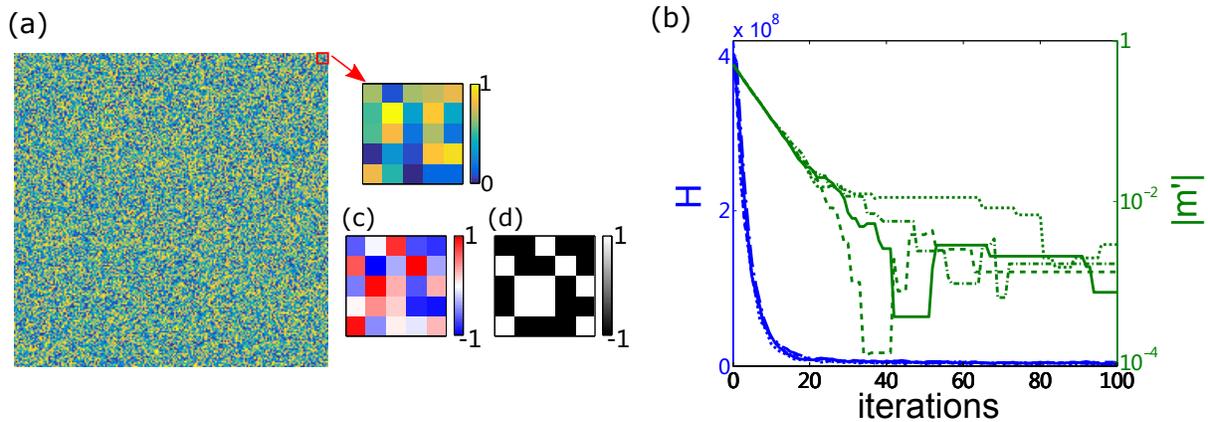}
\caption{\label{fig:3} Experimental results for the searching process of number-partitioning problem. (a) The set $\Xi =\{\xi_j\}$ containing $40000$ elements, where each $\xi_j$ is chosen randomly from $(0,1]$ with uniform probability. (b) Evolutions of the Hamiltonian $H$ and amplitude of the gauge-transformed magnetization $\left| {{m}'} \right|$ during the iterations. Four independent trials are conducted with uniform initial spin configurations. For clear visualization, (c) and (d) respectively presents only a part of the final configurations of the gauge-transformed spins $\left\{ {\sigma'_j}^{z} \right\}$ on SLM and the original spins $\left\{ {{\sigma }_{j}} \right\}$, which correspond to the red square box in (a). }
\end{figure*}

As an example, we demonstrate the searching process with the number-partitioning problem of a set $\Xi =\{\xi_j\}$ with $N=40000$ elements. The set $\Xi$ is randomly generated that each element $\xi_j$ is a real number randomly chosen from $(0,1]$, as presented in Fig.~\ref{fig:3}(a) in the form of a 200-by-200 array. Here each spin is encoded by a macropixel with 2-by-2 pixels on SLM with the length of ${W}=16\mu \text{m}$. Since the beam size is much larger than that of the array, we assume that light illuminates each macropixel with a uniform amplitude.

In order to evaluate the correlation function $F$, we first numerically calculate the distribution function $g_c(\mathbf{u})$ through Eq.~(\ref{eq:3}). Given the interaction strength between any two spins $ G({\mathbf{k}})=-1$, Fig.~\ref{fig:2}(a) shows the calculated distribution function $g_c(\mathbf{u})$.  We note that $g_c(\mathbf{u})$ have both positive and negative values, which means such a case cannot be implemented by the target-intensity approach as Ref.~\cite{pierangeli2019large}. We also need to calibrate the intensity measurement such that the distribution function $g_c(\mathbf{u})$ has the same origin as the intensity distribution $I(\mathbf{u})$. In order to reduce the impact of optical aberrations, we measure the intensity distribution on CCD plane [Fig.~\ref{fig:2}(b)] by setting the SLM with uniform phase modulation. Therefore, the origin of $I(\mathbf{u})$ is marked at the maximal intensity through the numerical fitting.

Next we start with the initial state that all spins are uniformly distributed as $\sigma_j=1$ and update the spin configuration for ground state search.  Here we utilize the Markov chain Monte Carlo algorithm, where $\sigma_j$s are tentatively updated during each iteration, and the gauge-transformed spins ${\sigma'_j}^z=\xi_j\sigma_j$ are encoded on SLM following Eq. (\ref{eq:1}). Then we measure the intensity $I(\mathbf{u})$ on the CCD and evaluate the system Hamiltonian through the correlation function $F$ as Eq.~(\ref{eq:2}). The updated spin configuration is accepted only when the Hamiltonian decreases.

Figure \ref{fig:3}(b) shows the evolution of the Hamiltonian $H$ and the amplitude of the gauge-transformed magnetization $\left| {{m}'} \right|$ during the ground-state-search process. In the experiment, four independent trials are conducted with the initial state that all spins are uniformly distributed. For all the four cases, the Hamiltonian $H$ and the magnetization $\left| {{m}'} \right|$ decrease rapidly at the beginning of the iteration, because the initial spin configuration strongly deviates from the ground state. As the number of iterations increases, the Hamiltonian tends to be stable, while the $\left| {{m}'} \right|$ starts to fluctuate. We attribute it to the too weak intensity distribution $I(\mathbf{u})$, which is strongly affected by the noise during the measurement. As a result, the spin configuration may be incorrectly updated with the distribution resulting in a larger $\left| {{m}'} \right|$. The accuracy can be improved by using a wide dynamic-range detector for intensity measurement, or by adjusting the input light intensity in real time within a suitable range.

Here for clear visualization, corresponding to the red square in Fig.~\ref{fig:3}(a), Figs. \ref{fig:3}(c) and (d) present a part of the final configurations of the gauge-transformed spins $\mathbf{{S}'}=\left\{ {\sigma'_j}^{z} \right\}$ and the resulting spin configuration $\left\{ {{\sigma }_{j}} \right\}$, respectively. Overall, for all these four trials, $\left| {{m}'} \right|$ reaches lower than $1.7\times 10^{-3}$ within $100$ iterations. Thus during the ground state search, the magnetization $\left| {{m}'} \right|$ decreases by nearly three orders of magnitude, which indicates the validity of the gauge transformation method for antiferromagnetic model.

\begin{figure}[htb]
\includegraphics{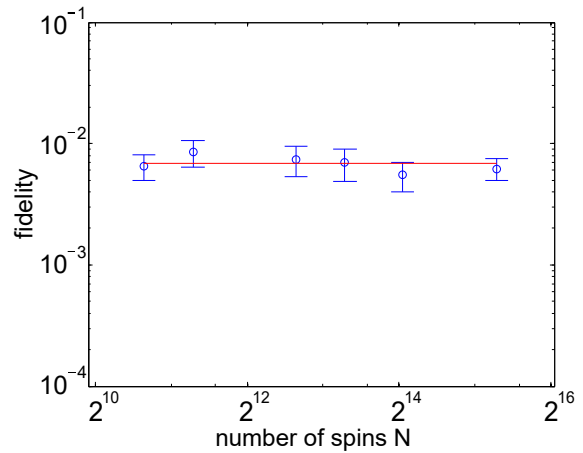}
\caption{\label{fig:4}The computing fidelities for number sets with sizes of $N=1600, 2500, 6400, 10000, 16900, 40000.$}
\end{figure}

To evaluate the scalability of the ground state search for number-partitioning problems, we define the computing fidelity as $\left| \frac{\sum_1-\sum_2}{\sum_1+\sum_2} \right|$, which is expected to be as small as possible like $\left| {{m}'} \right|$. We investigate its performance as a function of the size $N$ of the number set and perform experiments for sizes $N$ varying from 1600 to 40000. For each $N$, we perform 10 independent experimental trials with different sets $\Xi=\{\xi_j\}$, and then the averaged computing fidelity is obtained by measuring the final states after $1000$ iterations. From the results in Fig.~\ref{fig:4}, we observe that fidelity remains within $6.9\times 10^{-3}$, demonstrating that the experimental setup works effectively for large-scale number sets. The good scalability of SPIM with gauge transformation on number-partitioning problems is inherited in the parallelism of the optical analog signal processing in the spatial domain.

\section{Conclusion and Discussion}

We propose to implement antiferromagnetic model in SPIM. By gauge transformation, an antiferromagnetic Hamiltonian can be evaluated through the correlation between the distribution function and the measured optical intensity with SPIM. To improve the processing speed of the system, the ultrafast SLM and CCD at gigahertz rates with the most recent technologies \cite{chen2020ultra,park2021all} is helpful and practical. Also, the computing accuracy can be improved with a more sensitive CCD camera. We note that our proposed method can be applied to the ground-state-search process, e.g., adiabatic evolution and simulated annealing algorithms \cite{farhi2001quantum,kirkpatrick1983optimization}.

In summary, we optically demonstrate the ground-state-search process of an antiferromagnetic Mattis model with thousands of spins, as well as the number-partitioning problem. With the improved accuracy resulting from gauge transformation, we successfully reduce the total magnetization strength of the gauge-transformed spins $\left| {{m}'} \right|$ by nearly three orders of magnitude. Thus for practical applications in modeling statistical systems and studying combinatorial optimization problems, such an optoelectronic
computing exhibits great programmability and scalability in large-scale systems.

\section{Acknowledgement}
The authors acknowledge funding through the National Natural Science Foundation of China (NSFC Grants Nos. 91850108 and 61675179), the National Key Research and Development
Program of China (Grant No. 2017YFA0205700), the Open Foundation of the State Key Laboratory of Modern Optical Instrumentation, and the Open Research Program of Key Laboratory of 3D Micro/Nano Fabrication and Characterization of Zhejiang Province.


The authors declare no conflict of interest.



%

\end{document}